\documentclass{aa}
\usepackage{natbib}
\bibpunct{(}{)}{;}{a}{}{,}

\usepackage{graphicx}
\usepackage[varg]{txfonts}
\def\totdif#1#2{\frac{{\rm d}#1}{{\rm d}#2}}

\begin{document} 

\title{On the computation of eigenfrequencies for equilibrium models including turbulent pressure}
\subtitle{}

\author{T. Sonoi\inst{1}
  \and
  K. Belkacem\inst{1}
  \and
  M. -A. Dupret\inst{2}
  \and
  R. Samadi\inst{1}
  \and
  H.-G. Ludwig\inst{3,4}
  \and
  E. Caffau\inst{4}
  \and
  B. Mosser\inst{1}
}
%\runningauthor{}

\institute{LESIA, Observatoire de Paris, PSL Research University, CNRS, Universit\'e Pierre et Marie Curie,
  Universit\'e Denis Diderot, 92195 Meudon, France
  \and
  Institut d'Astrophysique et de G\'eophysique, Universit\'e de Li\`ege, All\'ee du 6 Ao\^ut 17, 4000, Li\`ege, Belgium
  \and
  Zentrum f\"ur Astronomie der Universit\"at Heidelberg, Landessternwarte, K\"onigstuhl 12, D-69117 Heidelberg, Germany
  \and
  GEPI, Observatoire de Paris, PSL Research University, CNRS, Universit\'e Denis Diderot, Sorbonne Paris Cit\'e, 5 Place Jules Janssen, 92195 Meudon, France
}

\offprints{\tt takafumi.sonoi@obspm.fr}
\date{\today}

% \abstract{}{}{}{}{} 
% 5 {} token are mandatory
 
\abstract
    {The space-borne missions CoRoT and 
%%%{\it Kepler} \LEt{ remove italics.}
Kepler 
have provided a wealth of highly accurate data. However, our inability to properly model the upper-most region of solar-like stars prevents us from making the best of these observations. This problem is called ``surface effect'' and a key ingredient to solve it is turbulent pressure for the computation of both the equilibrium models and the oscillations. While 3D hydrodynamic simulations help to include properly the turbulent pressure in the equilibrium models, the way this surface effect is included in the computation of stellar oscillations is still subject to uncertainties.}
    {We aim at determining how to properly include the effect of turbulent pressure and its Lagrangian perturbation in the adiabatic computation of the oscillations. We also discuss the validity of the 
%%%{\bf [NOTE: we would like to remove the part 'approximations introduced by Rosenthal et al. (1995, 1999)' to avoid using the reference.]}
gas-gamma model and reduced gamma model approximations, which have been used to compute adiabatic oscillations of equilibrium models including turbulent pressure.}
    {We use a patched model of the Sun with an inner part constructed by a 1D stellar evolution code (CESTAM) and an outer part by the 3D hydrodynamical code (CO$^5$BOLD). Then, the adiabatic oscillations are computed using the ADIPLS code for the gas-gamma and reduced gamma model approximations and with the MAD code imposing the adiabatic condition on an existing time-dependent convection formalism. Finally, all those results are compared to the observed solar frequencies.}
    {We show that the computation of the oscillations using the time-dependent convection formalism in the adiabatic limit improves significantly the agreement with the observed frequencies compared to the gas-gamma and reduced gamma model approximations. Of the components of the perturbation of the turbulent pressure, the perturbation of the density and advection term is found to contribute most to the frequency shift.}
    {The turbulent pressure is certainly the dominant factor responsible for the surface effects. Its inclusion into the equilibrium models is thus necessary but not sufficient. Indeed, the perturbation of the turbulent pressure must be properly taken into account for computing adiabatic oscillation frequencies. We propose a formalism to evaluate the frequency shift due to the inclusion of the term with the turbulent pressure perturbation in the variational principle in order to extrapolate our result to other stars at various evolutionary stages. Although this work is limited to adiabatic oscillations and the inclusion of the turbulent pressure, future works will have to account for the nonadiabatic effect and convective backwarming.}


   \maketitle
%
%________________________________________________________________

\section{Introduction}

As shown by the space missions CoRoT \citep{Baglin06cosp, Baglin06ESASP, Michel08} and 
%%%{\it Kepler\LEt{ remove italics.} } 
Kepler
\citep{Borucki10, Bedding10, Chaplin11}, solar-like oscillations are ubiquitous to low-mass stars from the main-sequence to the red giant branch. They have been widely used to infer the internal structure of those stars and have permitted us to dramatically improve our knowledge as well as to put stringent constraints on stellar structure and evolution \citep[e.g.][]{Chaplin13}.

However, there are still some fundamental difficulties to overcome so as to exploit the full potential of the asteroseismic observations. Surface effects are likely to be the most important. This generic term names the systematic differences between the observed and model frequencies due to our deficient physical description of the upper-most layers of solar-like stars \cite[e.g.][]{JCD2016}. One of the key ingredients of those surface effects is the turbulent pressure. In standard models of stellar equilibrium structure and oscillations, it is generally neglected because its modelling is difficult. Nevertheless, it is a key factor to obtain accurate frequencies of stellar models and particularly for $p$ modes that are very sensitive to the surface layers. The crucial role of turbulent pressure in computing stellar oscillations has been emphasized in many studies \citep{Brown84, Zhugzhda94, Schlattl97, Petrovay07, Houdek10}. More recently, analyses of surface effects have been carried out using 3D hydrodynamical models \citep{Stein91, Rosenthal95, Rosenthal99, Yang07, Piau14, Bhattacharya15, Sonoi15, Magic16, Ball16, Houdek17, Trampedach17}, because these models provide a realistic description of the equilibrium structure including the turbulent pressure.

\begin{table*}
  \caption{Characteristics of the patched model (PM)}
  \label{tab:model}
  \begin{tabular}{ccccccc}\hline\hline
    $T_{\rm eff}$ [K] & $\log\, g$ [g cm$^{-2}$] & $T_{\rm b}$ [K] & $p_{\rm b}$ [g cm$^{-1}$ s$^{-2}$] & $M$ [$M_\odot$] & Age [Gyr] & $\alpha$ \\ \hline
    5775 & 4.44 & $1.53\times 10^4$ & $3.66\times 10^6$  & 1.01           & 4.61      & 1.65 \\ \hline
  \end{tabular}
\end{table*}

However, when turbulent pressure is included in the equilibrium model, the computation of the related stellar oscillations becomes tricky and we have to care about the possible inconsistency between the oscillation formalism and the equilibrium models. To consider this problem, \cite{Rosenthal95, Rosenthal99} proposed two approximations, the gas-gamma model (GGM), for which the Lagrangian perturbation of the turbulent pressure equals to the perturbation of the gas pressure, and the reduced gamma model (RGM), for which the Lagrangian perturbation of the turbulent pressure vanishes. They have shown that the GGM frequencies better reproduce the observed frequencies compared to ones obtained with the RGM assumption. Their result implies that it is important to take the perturbation of the turbulent pressure into account in order to obtain accurate frequencies. However, the GGM assumption does not rely on a convincing principle and deserves more investigation.

In this work, we consider the computation of adiabatic oscillations for an equilibrium model including turbulent pressure. To do so, we use a time-dependent convection (TDC) formalism that enables us to account for the perturbation of turbulent pressure. We also discuss the validity of the GGM approximation. We use the TDC formalism developed by \cite{Grigahcene05}, which originates from the work of \cite{Unno67} and was generalized for nonradial oscillations by \cite{Gabriel75}. This formalism has been so far adopted for the computation of the full nonadiabatic oscillations in order to explain the excitation of the classical pulsators \cite[e.g.][]{Dupret05, Dupret08}, or to fit to the damping rates of the solar-like oscillations \cite[e.g.][]{Dupret06b, Belkacem12, Grosjean14}. \cite{Dupret06} developed it for treating the nonlocal convection. For our purpose, we will impose the adiabatic condition on this formalism to see the validity of the GGM approximation. Moreover, such an approach allows us to consider the effect of the turbulent pressure separately from the nonadiabatic effect, which is also expected to affect eigenfrequencies \citep{Houdek10}.

The paper is organized as follows: Section \ref{sec:modeleigen} introduces how to compute eigenfrequencies with turbulent pressure. Section \ref{sec:cause} discusses the dominant causes of the frequency shift due to the perturbation of the turbulent pressure. Section \ref{sec:conclusion} gives the conclusion.

\section{Modelling eigenfrequencies with turbulent pressure}
\label{sec:modeleigen}
\subsection{Equilibrium model}
\label{sec:equil}

We use the solar ``patched'' model (PM) described in \cite{Samadi07} and \cite{Sonoi15}. The inner part of this model was constructed using the 1D stellar evolution code CESTAM \citep{Marques13} while the near-surface layers have been obtained using temporal and horizontal averages of the 3D hydrodynamical simulation by the CO$^5$BOLD \citep{Freytag12} code with the CIFIST grid \citep{Ludwig09}. The turbulent pressure is thus included only in the 3D upper layers. The matching between the inner and outer layers have been computed through an optimization of the 1D model with a Lenvenberg-Marquardt algorithm. The constraints for the optimization are the effective temperature ($T_{\rm eff}$) of the 3D model, the gravity acceleration at the photosphere ($g$), and the temperature at the bottom of the 3D model ($T_{\rm b}$). As for the last one, the temperature at the level having the same total pressure with the bottom of the 3D model ($p_{\rm b}$) is matched with $T_{\rm b}$. The free parameters are the stellar age, total mass ($M$), and mixing length parameter ($\alpha$). The resulting values are provided in Table \ref{tab:model}. Our matching point is deep enough since, at the bottom of our 3D model, the fraction of the turbulent pressure to the total pressure is small enough ($\simeq 0.014$) that it does not affect frequencies of acoustic modes, of which amplitude is concentrated in the upper layers.

 PM is constructed by replacing the outer part of the optimized 1D model, which we call ``unpatched'' model (UPM), with the averaged 3D model. The additional support by turbulent pressure modifies the hydrostatic equilibrium so that, at the photosphere, the radius of PM is larger than UPM by about 0.02\% \cite[see also Table 2 and Fig. 2 in][]{Sonoi15}.

\subsection{Computation of adiabatic oscillations: the gas-gamma and reduced gamma approximations}
\label{sec:ad}

Following the work of \cite{Rosenthal95, Rosenthal99}, two approximations can be adopted to account for the turbulent pressure in the equilibrium model, namely the gas-gamma model (hereafter GGM) and the reduced gamma model (hereafter RGM).

The GGM assumes that the relative Lagrangian perturbation of turbulent pressure equals the relative Lagrangian perturbation of thermal pressure, which is the sum of gas and radiation pressures, and hence is equal to that of the total pressure,
\begin{eqnarray}
  \frac{\delta p_{\rm turb}}{p_{\rm turb}}\simeq
  \frac{\delta p_{\rm tot}}{p_{\rm tot}}
  \simeq\frac{\delta p_{\rm th}}{p_{\rm th}}=\Gamma_1\frac{\delta\rho}{\rho},
  \label{eq:GGM}
\end{eqnarray}
where $\delta$ denotes the Lagrangian perturbation, $p_{\rm turb}$ is the turbulent pressure, $p_{\rm th}$ is the thermal pressure, and $p_{\rm tot}(=p_{\rm th}+p_{\rm turb})$ is the total pressure.

The RGM approximation introduces the reduced $\Gamma_1$, defined as $\Gamma^r_1\equiv (p_{\rm th}/p_{\rm tot})\Gamma_1$. In this approximation, the Lagrangian perturbation of turbulent pressure is neglected:
\begin{eqnarray}
  \frac{\delta p_{\rm turb}}{p_{\rm turb}}=0.
\end{eqnarray}
We have thus
\begin{eqnarray}
  \frac{\delta p_{\rm tot}}{p_{\rm tot}}=\frac{p_{\rm th}}{p_{\rm tot}}\frac{\delta p_{\rm th}}{p_{\rm th}}
  =\Gamma^r_1\frac{\delta\rho}{\rho}.
\end{eqnarray}

\begin{figure}
  \centering
  \includegraphics[width=\hsize]{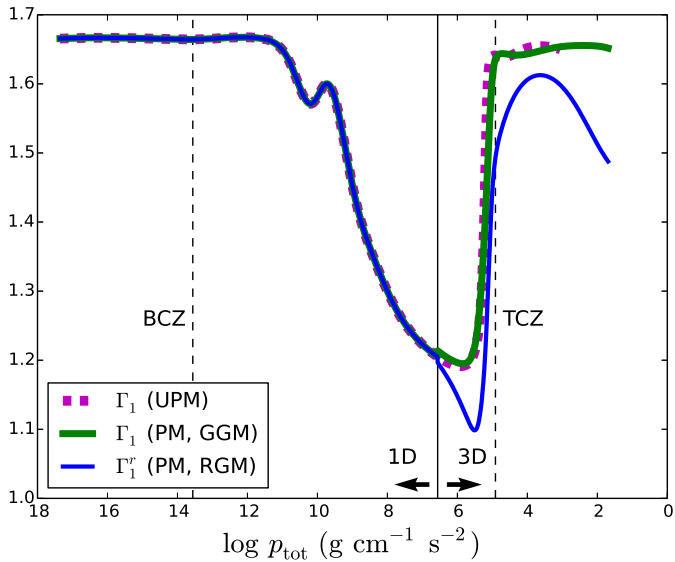}
  \caption{Adiabatic exponent as a function of total pressure in UPM (dashed magenta line) and PM (solid green and blue lines). For PM, ones for the GGM ($\Gamma_1$, green) and for the RGM ($\Gamma^r_1$, blue) are shown as functions of the total pressure. The vertical solid line indicates the matching point between 1D and 3D models. The vertical dashed lines indicate the bottom and top of the convection zone determined by the Schwarzschild criterion, labelled as BCZ and TCZ, respectively.}
  \label{fig:g1}
\end{figure}

Figure \ref{fig:g1} shows the adiabatic exponent in UPM and PM. We note
%%%\LEt{A and A does not allow direct instructions to the reader.} 
that there is no difference between GGM and RGM for UPM, which does not include any turbulent pressure. In the 3D layer of PM, $\Gamma^r_1$ has lower values than $\Gamma_1$ as a consequence of the presence of turbulent pressure.

\begin{figure}
  \centering
  \includegraphics[width=\hsize]{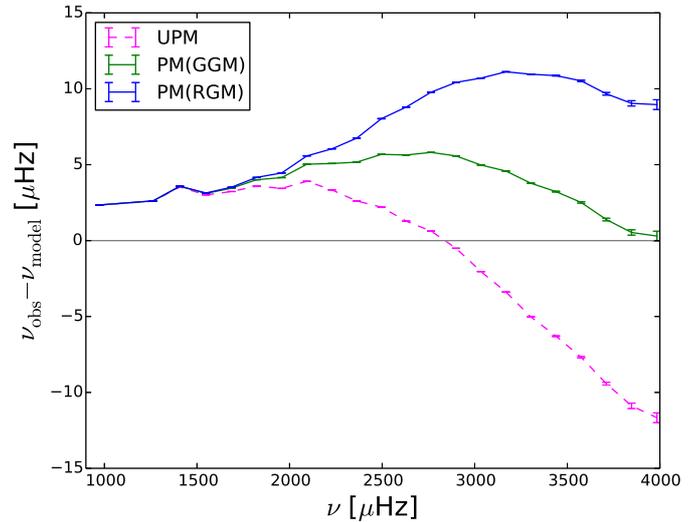}
  \caption{Difference between the model and observed frequencies \citep{Broomhall09} for the radial modes. The error bars stem from the observation. The magenta dashed line is for UPM, and the blue and green  solid lines are for PM with the RGM and GGM, respectively. The model frequencies are computed by ADIPLS.}
  \label{fig:freqdiff}
\end{figure}

Figure \ref{fig:freqdiff} shows the difference between the observed frequencies as given by \cite{Broomhall09} and the computed frequencies obtained using the PM and UPM described in Sect. \ref{sec:equil}. The larger radius of PM makes frequencies lower than for UPM. Then, the value of $(\nu_{\rm obs}-\nu_{\rm model})$ is higher for PM. On the other hand, the GGM treatment, namely including the perturbation of the turbulent pressure, oppositely increases the frequencies, and reduces the deviation of the RGM frequencies from the observed frequencies. As a result, the GGM frequencies are in better agreement with the observation than the RGM ones. This result confirms the result of \cite{Rosenthal99}.

The deviation of the GGM frequencies from the observation is at most $\sim 6\mu$Hz in our analysis. This is of the same order but a little larger than those of the other analyses using the other 3D hydrodynamical models ($\sim 4\mu$Hz in \citeauthor{Rosenthal99} \citeyear{Rosenthal99} and \citeauthor{Magic16} \citeyear{Magic16}, and $\sim 3\mu$Hz in \citeauthor{Ball16} \citeyear{Ball16}). The deviation of the RGM frequencies is at most $\sim 10\mu$Hz, similarly to \cite{Houdek17}.

Finally, we note that the GGM and RGM approximations are easily implemented in an adiabatic oscillation code. However, the underlying assumptions are rather crude and deserve more attention. Particularly, the perturbation of the turbulent pressure should be out of phase with that of the gas pressure and density \citep{Houdek00, Houdek17}. Then, computations only with the real part of the eigenfrequency such as those performed by the ADIPLS code are not valid. Therefore, one has to provide a modelling of the perturbation of the turbulent pressure and this is permitted by using a time-dependent modelling of convection (TDC) as provided in the following section. Indeed, the phase lag between the turbulent pressure and the other variables takes place in computation with TDC.

\subsection{Computation of adiabatic oscillations: the TDC treatment for nonlocal convection}
\label{sec:adtdc}

In the following, we adopt the TDC formalism developed by \cite{Grigahcene05} and \cite{Dupret06} to compute frequencies of PM. While it is usually used to compute non-adiabatic oscillations, we consider the limit of adiabatic oscillations by setting $\delta s=0$, where $s$ is the specific entropy. Such an approach allows us to properly consider both the effects of turbulent pressure on the equilibrium structure and of its perturbation in the adiabatic limit. Moreover, this clarifies the individual contribution of turbulent pressure to the surface effects separately from the nonadiabatic effects.

As mentioned at the end of Sect. \ref{sec:ad}, the phase lag occurs between the perturbation of the turbulent pressure and the other variables when we adopt a TDC formalism. The phase lag leads to excitation or damping of oscillation amplitude. Namely, the eigenfrequencies of the oscillation become complex. In this work, however, we only pay attention to the real part of the eigenfrequencies, since we need nonadiabatic treatment to exactly investigate the damping rates. We are aware that the nonadiabatic effects would be important not only for the damping rates, but also for the oscillation frequencies. This is however out of the scope of the present article and will be considered in a following
work.

Therefore, we start by considering the expression of the perturbation of the turbulent pressure,
\begin{eqnarray}
  \frac{\delta p_{\rm turb,l}}{p_{\rm turb,l}}=\frac{\delta\rho}{\rho}+2\frac{\overline{V_r\delta V_r}}{\overline{V^2_r}},
  \label{eq:dpturb}
\end{eqnarray}
where $p_{\rm turb,l}$ is the turbulent pressure as obtained in the framework of a local theory of convection and $V_r$ is the radial component of the convective velocity. The overbar indicates averaging in the coarse grain, which is much larger than most 
%%%of the \LEt{ or, if you are speaking more generally, most convective eddies.} 
%%%{\bf [NOTE: we would like to choose 'most convective eddies']}
convective eddies
but much smaller than the scale of the oscillation wavelength. To go further, we consider the perturbation of the radial convective velocity \citep[Eq. \ref{eq:dVr_nad} or Eq. 21 in ][]{Dupret06} in the adiabatic limit, but for the sake of simplicity, we limit ourselves to the case of radial oscillations ($\ell=0$). This gives
\begin{eqnarray}
  \frac{\overline{V_r\delta V_r}}{\overline{V^2_r}}&=&\frac{1}{B+[(i\Omega+\beta)\sigma\tau_c+1]D} 
  \cdot\left\{-\frac{\delta c_p}{c_p}-\frac{\delta Q}{Q} - \frac{\delta\rho}{\rho} \right.\nonumber\\
  &&\left.+\totdif{\delta p_{\rm tot}}{p_{\rm tot}} - (C+1)\totdif{\xi}{r}
  -\frac{A}{A+1}\frac{i\sigma\tau_c}{\Omega\Lambda}
  \left(\totdif{\xi}{r}
  +\frac{1}{A}\frac{\xi}{r}\right)\right.\nonumber\\
  &&\left.-\omega_R\tau_cD\frac{\delta\omega_R}{\omega_R}+(D+1)\frac{\delta l}{l}\right\},
  \label{eq:dVr}
\end{eqnarray}
with
\begin{eqnarray}
  B&=&\frac{i\sigma\tau_c+\Omega\Lambda}{\Omega\Lambda}, \\
  C&=&\frac{\omega_R\tau_c+1}{(i\Omega+\beta)\sigma\tau_c+\omega_R\tau_c+1}, \label{eq:C} \\
  D&=&\frac{1}{(i\Omega+\beta)\sigma\tau_c+\omega_R\tau_c+1}, \label{eq:D}
\end{eqnarray}
where $\sigma$ $(\equiv 2\pi\nu)$ is the oscillation frequency in unit of rad s$^{-1}$, $\tau_c$ is the convective timescale, $c_p$ is the specific heat capacity at constant pressure, $\rho$ is the density, $\xi$ is the displacement, $\omega_R$ is the inverse of the radiative cooling timescale of convection eddies, $l$ is the mixing length defined by Eq. (\ref{eq:lmix}), and $Q[\equiv -(\partial\ln\rho/\partial\ln T)_{p_{\rm th}}]$ is the volume expanding rate.

The free parameters
$\beta$ and $\Omega$ 
%%%\LEt{ Please avoid beginning a sentence with a symbol. Suggestion: The free parameter...are related....} 
are 
%%%free parameters 
related to the closure of the TDC theory. 
%%%\LEt{ see note 7.} 
The parameter
$\beta$ is a complex value and is introduced in Eq. (\ref{eq:beta}). 
%%%\LEt{ as above.} 
The parameter
$\Omega$ is an adjusting function introduced in the closure terms of the momentum and energy equations for the convective fluctuations (Eqs. \ref{eq:movement_mod} and \ref{eq:energy_mod}). For stationary convection, it has the same meaning as in the formalism of \cite{Canuto91}. This quantity is determined by matching with the results given by the 3D simulation using Eqs. (\ref{eq:MLT_pturb}b), (\ref{eq:MLT2_nl}) and (\ref{eq:Fc_om}).

\begin{figure}
  \centering
  \includegraphics[width=\hsize]{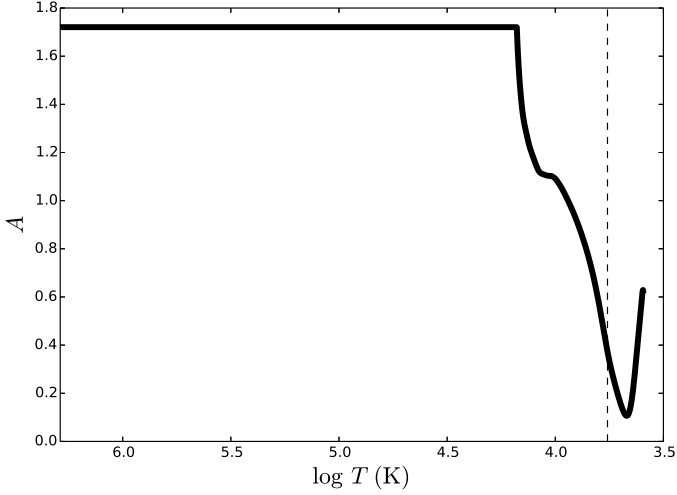}
  \caption{Anisotropy parameter $A$ as a function of temperature in log-scale in the range from the bottom of the convection zone to the top of the atmosphere for PM. The vertical line indicates the upper boundary of the convection zone determined by the Schwarzschild criterion.}
  \label{fig:A}
\end{figure}

The parameter $A$ stands for the anisotropy of the turbulence and is defined as
\begin{eqnarray}
  A=\frac{\overline{\rho V^2_r}}{\overline{\rho(V^2_\theta+V^2_\phi)}},
  \label{eq:A}
\end{eqnarray}
where $V_\theta$ and $V_\phi$ are the horizontal components of the convective velocity. In this work, this parameter is obtained directly from the 3D simulation. For the layers extracted from the 1D model, we fix the value as given at the bottom of the 3D simulation. This quantity is displayed in Fig. \ref{fig:A}.

\begin{figure}
  \centering
  \includegraphics[width=\hsize]{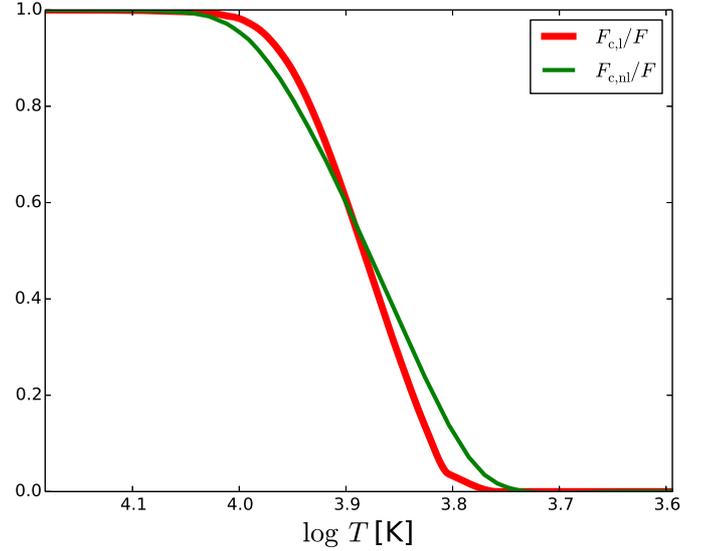}
  \caption{Temporally and horizontally averaged 3D convective flux, $F_{\rm c,nl}$, and its local counter part obtained by Eq. (\ref{eq:fcnl}), $F_{\rm c,l}$ for PM. The values are normalized by the total flux, $F$.}
  \label{fig:fcnl}
\end{figure}

For taking the non-locality into account, we adopt the approach of \cite{Spiegel63}. It consists in using an analogy with radiative transfer. The local values, as given by the 
%%%MLT\LEt{Please give full form of acronym on first usage.}, 
mixing length theory (MLT),
are considered as source terms and then the nonlocal values are obtained by performing an average, that is,
\begin{eqnarray}
  && p_{\rm turb,nl}(\zeta_0)=\int^{+\infty}_{-\infty}p_{\rm turb,l}{\rm e}^{-b|\zeta-\zeta_0|}{\rm d}\zeta, \\
  && F_{\rm c,nl}(\zeta_0)=\int^{+\infty}_{-\infty}F_{\rm c,l}{\rm e}^{-a|\zeta-\zeta_0|}{\rm d}\zeta,
\end{eqnarray}
where ${\rm d}\zeta={\rm d}r/H_p$, and $a$ and $b$ are free parameters as introduced by \cite{Balmforth92}. The temporally and horizontally averaged values of turbulent pressure and convective flux in the 3D model are substituted into $p_{\rm turb,nl}$ and $F_{\rm c,nl}$, respectively. 
%%%\LEt{ see note 7.}  
The quantities
$p_{\rm turb,l}$, $F_{\rm c,l}$ stand for their local counterparts. These equations can be recast by taking the second order derivative
\begin{eqnarray}
  && {\rm d}^2p_{\rm turb,nl}/{\rm d}\zeta^2=b^2(p_{\rm turb,nl}-p_{\rm turb,l}), \label{eq:ptnl} \\
  && {\rm d}^2F_{\rm c,nl}/{\rm d}\zeta^2=a^2(F_{\rm c,nl}-F_{\rm c,l}). \label{eq:fcnl}
\end{eqnarray}
Equations (\ref{eq:ptnl}) and (\ref{eq:fcnl}) are then used to infer the values of $a$ and $b$ as well as the local values of the turbulent pressure and convective flux from the 3D numerical simulation. In the overshooting region, the two local quantities, $p_{\rm turb,l}$ and $F_{\rm c,l}$, vanish, so that $a$ and $b$ are obtained by fitting an exponential function to the turbulent pressure and convective flux as given by the 3D simulation. From our model, we get $a=6.975$ and $b=1.697$. Subsequently, the local counterparts ($p_{\rm turb,l}$ and $F_{\rm c,l}$) are easily obtained by solving Eqs. (\ref{eq:ptnl}) and (\ref{eq:fcnl}) in the convective region (Fig. \ref{fig:fcnl} and top panel of Fig. \ref{fig:rept}).
%%%\LEt{ one sentence cannot constitute a paragraph alone. Please attach to preceding or following para as appropriate.} 
With the equations for stationary convection (Eqs. \ref{eq:MLT1}, \ref{eq:MLT_pturb}b and \ref{eq:Fc_om}), we can evaluate $\Omega$, $\alpha$ and $\Gamma$ as functions of the depth, where $\Gamma\,[=(\omega_R\tau_c)^{-1}]$ is the convective efficiency. 

The perturbations of the nonlocal turbulent pressure and convective flux ($\delta p_{\rm turb,nl}$ and $\delta F_{\rm c,nl}$) are obtained by solving the eigenvalue problem of the perturbed hydrodynamical equations of mean flow combined with the perturbed equations of (\ref{eq:ptnl}) and (\ref{eq:fcnl}). On the other hand, their local counterparts ($\delta p_{\rm turb,l}$ and $\delta F_{\rm c,l}$) can be evaluated with the linear combination of the eigenfunctions, which is given by Eqs. (\ref{eq:dpturb}) and (\ref{eq:dVr}) for $\delta p_{\rm turb,l}$, and Eq. (\ref{eq:dfc}) for $\delta F_{\rm c,l}$.

\subsection{Comparison of frequencies among GGM, RGM, and TDC}
\label{sec:compare_freq}

\begin{figure}
  \centering
  \includegraphics[width=\hsize]{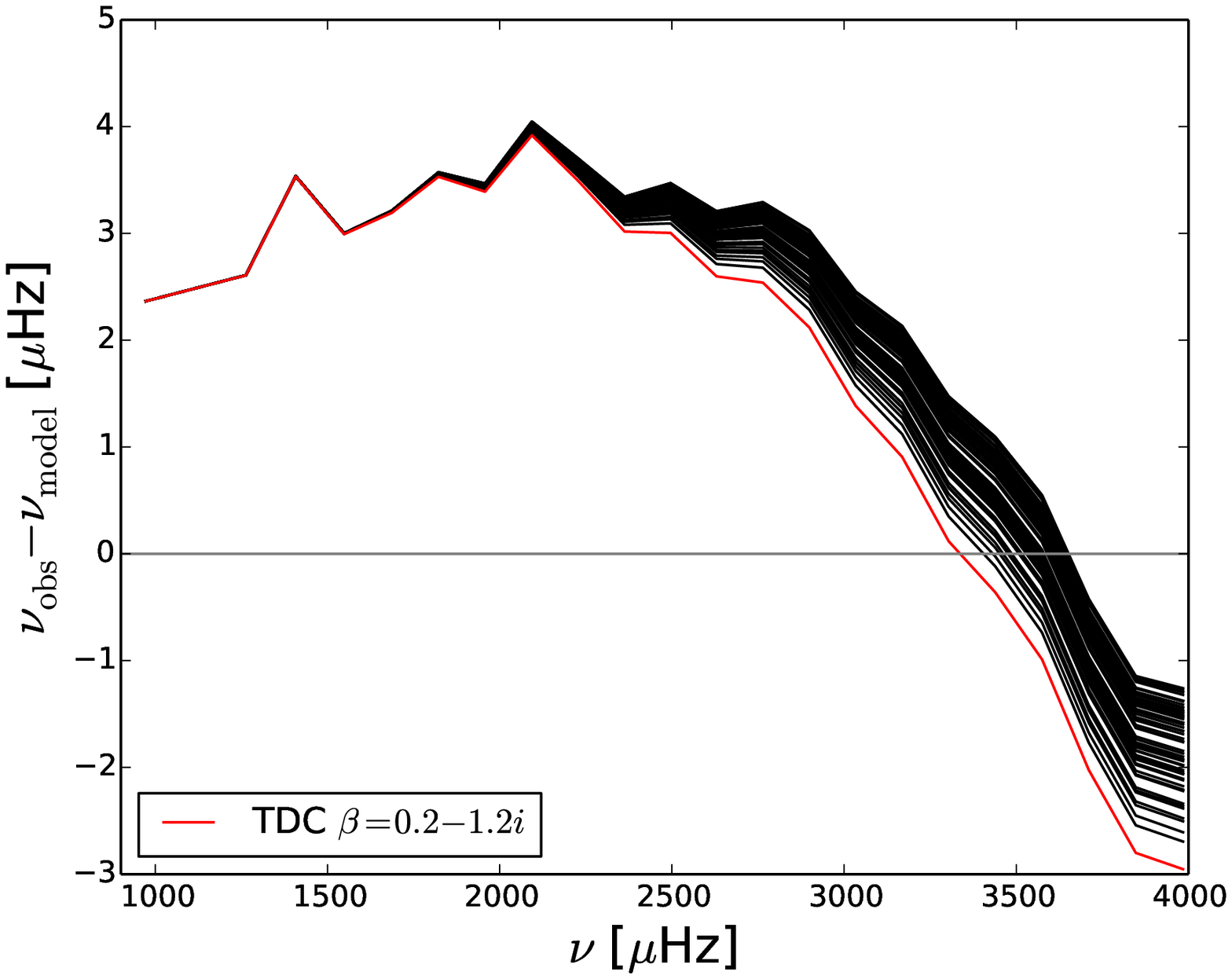}
  \includegraphics[width=\hsize]{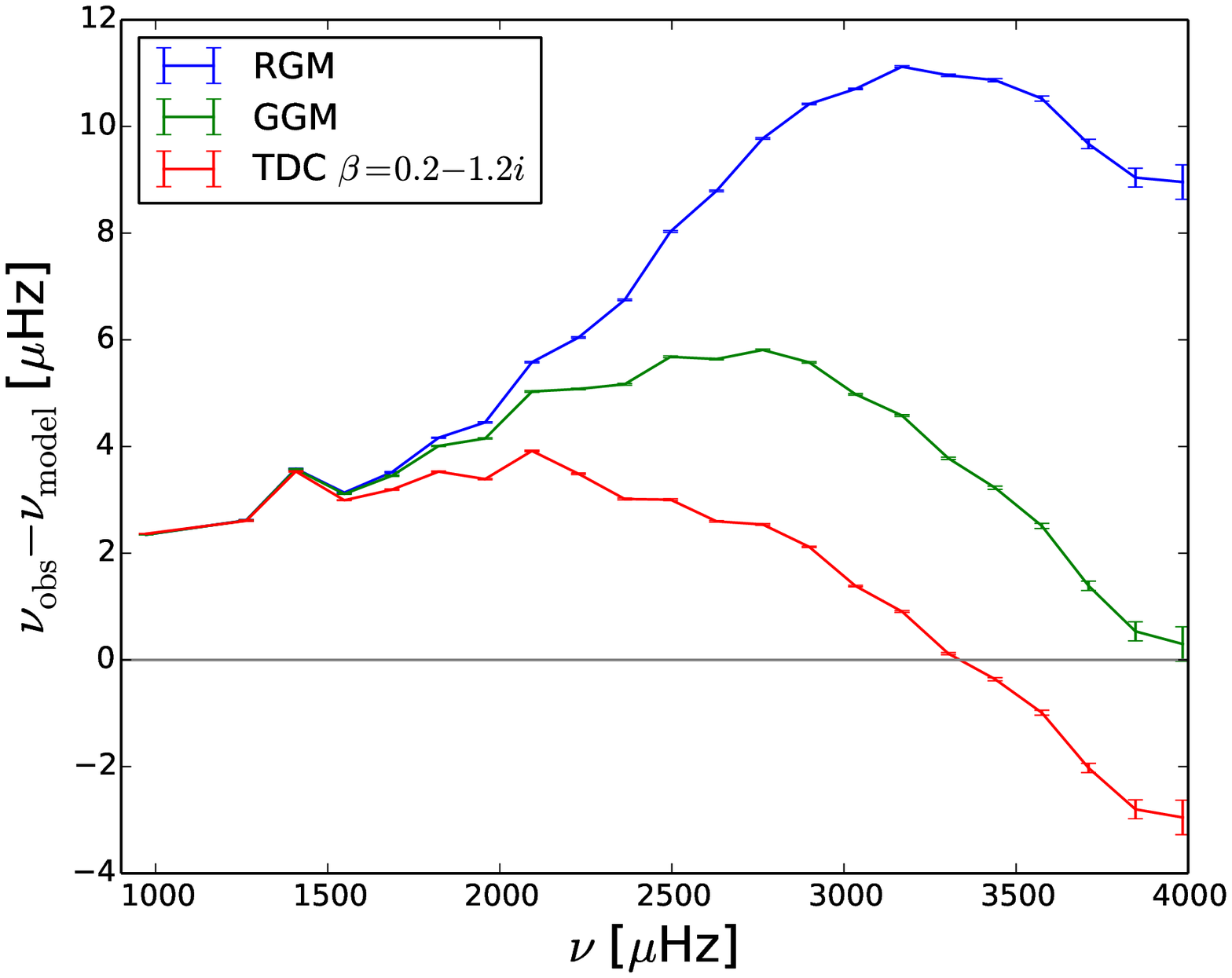}
  \caption{Same as Fig. \ref{fig:freqdiff}, but for the model frequencies computed with TDC for PM.
Top: the black lines are for different values of the TDC free parameter $\beta$. The real part of $\beta$ are ranged from 0.4 to 2.0, while the imaginary part from $-$2.0 to 2.0 at 0.4 intervals. The red line is for $\beta=0.2-1.2i$, which gives the smallest deviation from the observed frequencies. Bottom: comparison of the case of $\beta=0.2-1.2i$ with GGM and RGM for PM.}
  \label{fig:freqdiff_TDC}
\end{figure}

For the computation with the TDC formalism, we need to give the value of the free parameter $\beta$. The calibrated $\beta$ values have been of the order of unity in the previous studies \citep{Dupret05, Dupret06b, Dupret08, Belkacem11, Grosjean14}. In this work, the real part were ranged from 0.2 to 2.0, while the imaginary part from $-$2.0 to 2.0 at 0.2 intervals. The top panel of Fig. \ref{fig:freqdiff_TDC} shows the results with the different values of $\beta$ (the 
%%%{\it black\LEt{ \LEt{Please remove the italics here. A and A does not use italics for emphasis.}.} } 
black
lines, at 0.4 intervals in both the real and imaginary parts for visibility). Evaluating $\chi^2=\sum_n(\nu^n_{\rm model}-\nu^n_{\rm obs})^2$ for each $\beta$, we found that the value of $\beta=0.2-1.2i$ gives the smallest deviation from the observed frequencies (the 
%%%{\it red} \LEt{ italics.} 
red
line). 

The bottom panel compares the case of $\beta=0.2-1.2i$ with the GGM and RGM. First, compared to the GGM, the TDC treatment improves the agreement with the observations, particularly for the intermediate radial order modes. The deviation from the observed frequencies is at most $\sim$ 4$\mu$Hz. Although our analysis is adiabatic, it provides results of the same order as the nonadiabatic analysis of \cite{Houdek17}, who used another TDC formalism \citep{Gough77a,Gough77b} and PM with a 3D model of \cite{Trampedach13} and reported $\sim 3\mu$Hz deviation from the observed frequencies.

Secondly, the GGM frequencies are closer to the TDC ones than the RGM ones are. Although \cite{Rosenthal95,Rosenthal99} and our results in Sect. \ref{sec:ad} (Fig. \ref{fig:freqdiff}) show that the GGM reproduced the observations better than the RGM, this result implies that the GGM is superior to the RGM also from the theoretical point of view. Namely, it would be worth taking the perturbation of the turbulent pressure into account even for the adiabatic computations from both observational and theoretical viewpoints.

We note some difference between the TDC and GGM for $n\gtrsim 10$, while their frequencies are almost identical for the lower radial orders. This difference implies that the turbulent pressure perturbation is not as simple as provided by Eq. (\ref{eq:GGM}), and that the GGM cannot reproduce the influence of the turbulent pressure with enough precision. We discuss such effects in the following section.

\section{Contribution to the frequency shift introduced by turbulent pressure perturbation}
\label{sec:cause}

In Sect. \ref{sec:ad} and Fig. \ref{fig:freqdiff}, we have shown the frequency shift due to the elevation of the upper layer due to the turbulent pressure in the equilibrium model, comparing the PM and UPM. Here, we discuss the contribution to the frequency shift due to the perturbation of the turbulent pressure. First, we determine which region in the star contributes to the frequency shift (Sect. \ref{sec:zone_freq_shift}). As shown in Eqs. (\ref{eq:dpturb}) and (\ref{eq:dVr}), the perturbation of the turbulent pressure consists in different perturbative processes. Secondly, we determine which perturbative process in convection is dominant (Sect. \ref{sec:main_eff}).

\subsection{Contributing region to the frequency shift}
\label{sec:zone_freq_shift}

To see the contribution of the turbulent pressure perturbation, we adopt the variational principle. Multiplying $\xi_r^*$ in both sides of the equation of movement (\ref{eq:pulsmov}), using Eqs. (\ref{eq:pulscont}) and (\ref{eq:poisson}), integrating over the mass of the star and taking the real part, we obtain
\begin{eqnarray}
   \nu^2&=& \frac{1}{4\pi^2}\left(\int^M_0|\xi|^2{\rm d}m\right)^{-1}
    \int^M_0\left({\rm Re}\left[\frac{\delta\rho^*}{\rho}
      \frac{(\delta p_{\rm th}+\delta p_{\rm turb})}{\rho}\right]\right. \nonumber\\
    && \left.-2\frac{g}{r}|\xi|^2 
    +\frac{2A-1}{A}\frac{p_{\rm turb,l}}{\rho}
    {\rm Re}\left[\frac{\xi^*}{r}\totdif{\xi}{r}\right]\right){\rm d}m.
  \label{eq:vp}
\end{eqnarray}
Except for low-order modes, the terms in the second line of Eq. (\ref{eq:vp}) hardly contribute since $|\xi/r|\ll |d\xi/dr|$ and $p_{\rm turb}\ll p_{\rm th}$. Here, we discuss the term with the turbulent pressure perturbation, $\delta p_{\rm turb}$. We introduce
\begin{eqnarray}
  N_{\rm turb}(m)=
  \frac{1}{8\pi^2\nu}
  \left(\int^M_0|\xi|^2{\rm d}m'\right)^{-1}
  \int^m_0{\rm Re}\left[\frac{\delta\rho^*}{\rho}\frac{\delta p_{\rm turb}}{\rho}\right]
      {\rm d}m',
  \label{eq:N_turb}
\end{eqnarray} 
so that the integral to the surface, $N_{\rm turb}(m=M)$, represents the frequency shift to which the turbulent pressure perturbation contributes. To be exact, this term includes some part of the effect of the upper layer elevation, which appears in the equilibrium variables, $\rho$ and ${\rm d}m$. Nevertheless, it is useful to see the contribution of the turbulent pressure perturbation. Indeed, since $p_{\rm th}\gg p_{\rm turb}$ and hence |$\delta p_{\rm th}|\gg |\delta p_{\rm turb}|$, most part of the elevation effect is included in the term with $\delta p_{\rm th}$.

\begin{figure}
  \centering
  \includegraphics[width=\hsize]{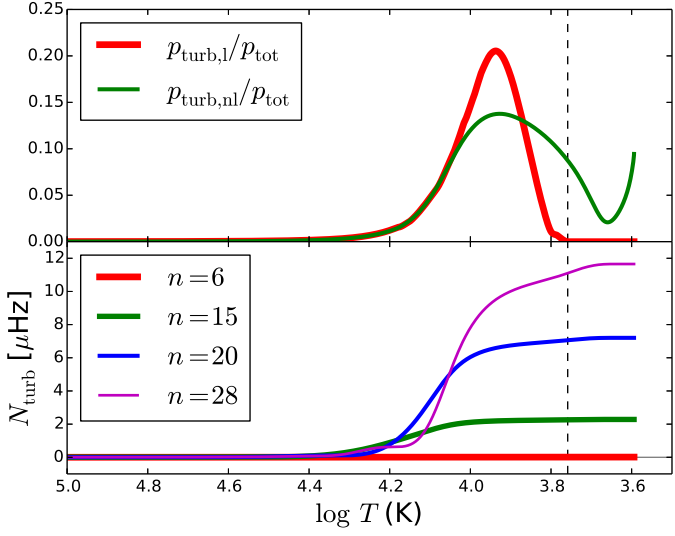}
  \caption{Top: Ratios of the local 
%%%({\it red\LEt{ italics.} }) 
(red)
and nonlocal turbulent pressures 
%%%({\it green\LEt{ italics.} }) 
(green)
to the total one. Bottom: cumulative contribution of the turbulent pressure perturbation to the eigenfrequency, $N_{\rm turb}$ (Eq. \ref{eq:N_turb}), for four radial modes with $\beta=0.2-1.2i$. The horizontal axis is the logarithm of temperature. The vertical dashed line indicates the upper boundary of the convection zone determined by the Schwarzschild criterion.}
  \label{fig:rept}
\end{figure}

The bottom panel of Fig. \ref{fig:rept} shows the profiles of $N_{\rm turb}$ for four radial modes. Here, the perturbation of the nonlocal turbulent pressure given by the MAD code is substituted into Eq. (\ref{eq:N_turb}). 
%%%$\LEt{ please avoid beginning a sentence with a symbol.} N_{\rm turb}$ 
The integral
$N_{\rm turb}$ increases mainly at $\log\,T\simeq$ 4.0--4.4, just below the peak of the $p_{\rm turb}/p_{\rm tot}$ ratio, shown in the top panel. By the way, it slightly increases even in the overshooting region above the boundary determined by the Schwarzschild criterion, since the nonlocal turbulent pressure contributes there. As the radial order $n$ increases, $N_{\rm turb}$ increases more substantially. The inertia gives the major contribution to this tendency. With the increasing radial order, the amplitude becomes confined in the near-surface region. Because of the low density of this region, the frequency becomes easier to shift. We discuss the dominant causes of the frequency shift in the following section.

\subsection{Dominant perturbative process}
\label{sec:main_eff}

\begin{figure}
  \centering
  \includegraphics[width=\hsize]{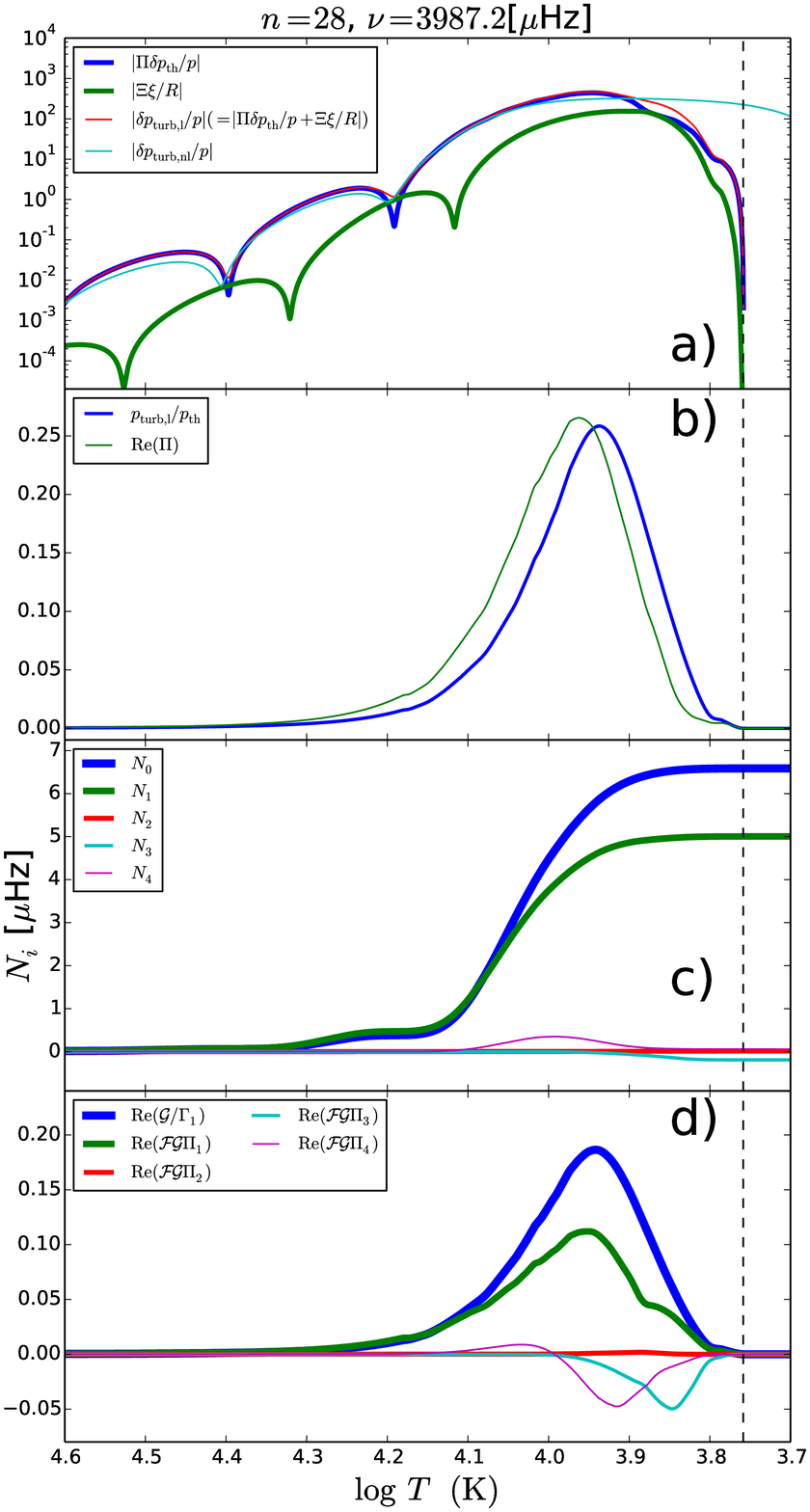}
  \caption{{\bf a)} absolute values of eigenfunctions for the $n=28$ mode obtained with the MAD code with $\beta=0.2-1.2i$; first and second terms in the right hand side of Eq. (\ref{eq:dpturb_local})
%%%({\it blue} 
(blue
and 
%%%{\it green},
green
%%%\LEt{ italics here and below.}  
respectively), perturbation of the local turbulent pressure given by Eq. (\ref{eq:dpturb_local}) 
%%%({\it red}), 
(red),
and of the nonlocal turbulent pressure 
%%%({\it cyan}).  
(cyan)
{\bf b)} real part of the coefficient for the perturbation of the thermal pressure in Eq. (\ref{eq:dpturb_local}), ${\rm Re}(\Pi)$, the ratio of the local turbulent pressure to the thermal one, $p_{\rm turb,l}/p_{\rm th}$. {\bf c)} cumulative contribution to the frequency shift of each decomposed component of $\Pi$ (Eq. \ref{eq:Pi}) defined by Eqs. (\ref{eq:N0}) and (\ref{eq:N1234}). {\bf d)} real parts of the decomposed components of $\Pi$ (Eq. \ref{eq:Pi}). The vertical dashed line indicates the upper boundary of the convection zone determined by the Schwarzschild criterion.}
  \label{fig:logT_vs_dPP_Pi}
\end{figure}

In the previous section, we have confirmed that the zone just below the peak of the $p_{\rm turb}/p_{\rm tot}$ ratio dominantly contributes to the frequency shift. Here we identify the respective contribution of the different processes to the total perturbation of turbulent pressure. For this purpose, we recast the expression of the perturbation of the local turbulent pressure (Eqs. \ref{eq:dpturb} and \ref{eq:dVr}). More precisely, we express it as the linear combination of the thermal pressure perturbation $\delta p_{\rm th}$ and the displacement $\xi$. The detailed procedure is described in Appendix \ref{app:transf_dpt}. Then, the perturbation of the local turbulent pressure (Eq. \ref{eq:dpturb}) is expressed as
\begin{eqnarray}
  \frac{\delta p_{\rm turb,l}}{p_{\rm tot}}=\Pi\frac{\delta p_{\rm th}}{p_{\rm tot}}+\Xi\frac{\xi}{R}
  \label{eq:dpturb_local}
\end{eqnarray}
with
\begin{eqnarray}
  &&\Pi={\cal G}\left[\frac{1}{\Gamma_1}+{\cal F}(\Pi_1+\Pi_2+\Pi_3+\Pi_4)\right], 
  \label{eq:Pi} \\
  &&\Xi={\cal FH}[\Xi_1+\Xi_2+\Xi_3+\Xi_4],
  \label{eq:Xi}
\end{eqnarray}
\begin{eqnarray}
  {\cal F}=\frac{2}{B+[(i\Omega+\beta)\sigma\tau_c+1]D},
  \label{eq:calF}
\end{eqnarray}
\begin{eqnarray}
  {\cal G}
  =\frac{p_{\rm turb,l}}{p_{\rm th}}
  \left[1-\frac{\cal F}{1+(\sigma\tau_c)^2}\frac{p_{\rm turb,l}}{p_{\rm tot}}
    (2\omega_R\tau_cD+1)(D+1)
    \right]^{-1},
\end{eqnarray}
and
\begin{eqnarray}
  {\cal H}
  &=&\frac{r}{R}\frac{p_{\rm turb,l}}{p_{\rm tot}} \nonumber \\
  &&\times\left[1-\frac{\cal F}{1+(\sigma\tau_c)^2}\frac{p_{\rm turb,l}}{p_{\rm tot}}
    (2\omega_R\tau_cD+1)(D+1)
    \right]^{-1},
  \label{eq:calH}
\end{eqnarray}
where the definitions of $\Pi_{1,2,3,4}$ and $\Xi_{1,2,3,4}$ are given by Eqs. (\ref{eq:pi1}) to (\ref{eq:xi4}). The 
%%%symbols 
%%%{\bf coefficients [NOTE: we would like to use 'coefficients' instead of 'symbols'.]}
coefficients
$\Pi_1$ and $\Xi_1$ correspond to the advection term in the equation of movement, $\Pi_2$ and $\Xi_2$ to the perturbation of the mixing length, $\Pi_3$ and $\Xi_3$ to the perturbation of the radiative cooling timescale of convection eddies, and $\Pi_4$ and $\Xi_4$ to the remaining parts.

Panel a) of Fig. \ref{fig:logT_vs_dPP_Pi} shows the absolute values of the eigenfunctions obtained with the MAD code. The perturbation of the local turbulent pressure reduces to zero toward the boundary determined by the Schwarzschild criterion 
%%%({\it red} \LEt{ italics here and below.} 
(red line). However, the perturbation of the nonlocal turbulent pressure has amplitude even in the overshooting zone due to the nonlocal effects expressed as Eq. (\ref{eq:ptnl}) 
%%%({\it cyan} 
(cyan line).

As shown in panel a), the second term of Eq. (\ref{eq:dpturb_local}) is negligible 
%%%({\it green} 
(green line). Then, Eq. (\ref{eq:dpturb_local}) simplifies to $\delta p_{\rm turb,l}/p_{\rm tot}\simeq\Pi\delta p_{\rm th}/p_{\rm tot}$. Using the adiabatic relation $\delta\rho/\rho=\delta p_{\rm th}/p_{\rm th}/\Gamma_1$, Eq. (\ref{eq:N_turb}) becomes
\begin{eqnarray}
  N_{\rm turb}(m)&\simeq&\frac{1}{8\pi^2\nu}
  \left(\int^M_0|\xi|^2{\rm d}m'\right)^{-1}\nonumber\\
  &&\times\int^m_0\frac{p_{\rm th}}{\Gamma_1\rho}
  \left|\frac{\delta p_{\rm th}}{p_{\rm th}}\right|^2
      {\rm Re}(\Pi)\,{\rm d}m'.
\end{eqnarray}
Although $\Pi$ is a complex number, we should pay attention to only its real part to discuss the frequency shift. Panel b) shows that the real part of $\Pi$ has a peak ($\log\,T\simeq 3.96$) located deeper than $p_{\rm turb,l}/p_{\rm th}$, which corresponds to the GGM approximation. However, their values are of the same order. It implies that the GGM treatment gives a good prediction to some extent.

Indeed, we can analytically understand that the GGM is valid in the bottom part of the convection zone. Since $\sigma\tau_c\gg 1\gg\omega_R\tau_c$ in such a part, we have
\begin{eqnarray}
  {\cal F}\rightarrow\frac{\Omega\Lambda}{i\sigma\tau_c},\;\;
  {\cal G}\rightarrow\frac{p_{\rm turb,l}}{p_{\rm th}},
\end{eqnarray}
and ${\cal G}/\Gamma_1$, which corresponds to the density perturbation in Eq. (\ref{eq:dpturb}), and ${\cal GF}\Pi_1$ are much larger than the other terms in Eq. (\ref{eq:Pi}). Therefore, we can derive
\begin{eqnarray}
  \frac{\delta p_{\rm turb,l}}{p_{\rm tot}}
  \rightarrow\frac{p_{\rm turb,l}}{p_{\rm th}}\frac{1}{\Gamma_1}
  \left(1+\frac{2A}{A+1}\right)\frac{\delta p_{\rm th}}{p_{\rm tot}}
  \sim\frac{p_{\rm turb,l}}{p_{\rm th}}\frac{\delta p_{\rm th}}{p_{\rm tot}},
  \label{eq:dpt_pi01}
\end{eqnarray}
which implies that the situation is close to the GGM (Eq. \ref{eq:GGM}) in the bottom part of the convection zone. 

To see the contribution of each component of $\Pi$ to the frequency shift, we introduce the variational principle like Eq. (\ref{eq:N_turb}):
\begin{eqnarray}
  N_0(m)=
  \frac{1}{8\pi^2\nu}\left(\int^M_0|\xi|^2{\rm d}m'\right)^{-1}
  \int^m_0\frac{p_{\rm th}}{\Gamma_1\rho}\left|\frac{\delta p_{\rm th}}{p_{\rm th}}\right|^2
  {\rm Re\left(\frac{\cal G}{\Gamma_1}\right)}\,{\rm d}m'
  \label{eq:N0}
\end{eqnarray}
and
\begin{eqnarray}
  N_i(m)&=&\frac{1}{8\pi^2\nu}
  \left(\int^M_0|\xi|^2{\rm d}m'\right)^{-1}\nonumber\\
  &&\times\int^m_0\frac{p_{\rm th}}{\Gamma_1\rho}
  \left|\frac{\delta p_{\rm th}}{p_{\rm th}}\right|^2
             {\rm Re}\left({\cal GF}\Pi_i\right){\rm d}m'
  \label{eq:N1234}
\end{eqnarray}
for $i=1,2,3,4$. Panel c) shows that the terms with ${\cal G}/\Gamma_1$ and $\Pi_1$ dominantly contribute to the frequency shift. Although Fig. \ref{fig:logT_vs_dPP_Pi} shows the case of $n=28$, the contributions of $\Pi_2$, $\Pi_3$ , and $\Pi_4$ are even more negligible for the other lower-order modes since the mode amplitude is distributed in the inner region. Panel d) shows that all the terms except $\Pi_2$, related to the perturbation of the mixing length, certainly contribute to the perturbation of the turbulent pressure in the top part of the convection zone. Particularly, the low convective efficiency, namely the low value of $\Gamma\,[=(\omega_R\tau_c)^{-1}]$, makes $\Pi_3$ contributive near to the upper boundary of the convection zone. However, the physical processes in the top part of the convection zone hardly contribute to the frequency shift. Then, Eq. (\ref{eq:N_turb}) would be written as
\begin{eqnarray}
  \Delta\nu_{\rm turb}&\simeq&\frac{1}{8\pi^2\nu}
  \left(\int^M_0|\xi|^2{\rm d}m\right)^{-1}\nonumber\\
  &&\times\int^M_0\frac{p_{\rm th}}{\Gamma_1\rho}\left|\frac{\delta p_{\rm th}}{p_{\rm th}}\right|^2
       {\rm Re}\left[{\cal G}\left(\frac{1}{\Gamma_1}+{\cal F}\Pi_1\right)\right]
       {\rm d}m.
       \label{eq:dnu_turb}
\end{eqnarray}
We note that we restricted our analysis to radial oscillations in this work. For nonradial oscillations, we should adopt Eq. (\ref{eq:dVr_nad}) instead of Eq. (\ref{eq:dVr}). Besides, Eqs. (\ref{eq:pulscont}) and (\ref{eq:poisson}) are no longer valid in the derivation. For high $\ell$ modes, the quantity $\ell(\ell+1)\xi_h$ may become important.

\section{Conclusion}
\label{sec:conclusion}

Previous studies \citep{Rosenthal95, Rosenthal99} have found that the frequencies obtained with the gas-gamma model (GGM) approximation better agree with the observations than those obtained with the reduced gamma model (RGM) approximation. This treatment is easy to adopt for computing the adiabatic oscillations of models including the turbulent pressure. However this crude approximation has no clear physical background. In this study, we computed the frequencies with a TDC formalism imposing the adiabatic condition. We found that the GGM provides closer frequencies to the TDC ones compared to the RGM. It implies that the GGM is superior to the RGM from not only observational but also theoretical viewpoints. Besides, the TDC computations reproduced the frequencies closer to the observation than did the GGM, regardless of the values of the free parameter $\beta$. Although our work is limited to the Sun, it is worth extrapolating our results obtained by the TDC to other stars. Using the variational principle, we found that the perturbation of the density and advection term mainly contribute to the frequency shift due to the perturbation of the turbulent pressure. Equation (\ref{eq:dnu_turb}) can be then used to evaluate the frequency shift for adiabatic radial oscillations.

As discussed in previous studies \cite[e.g.][]{Brown84,Rosenthal99,Sonoi15}, the turbulent pressure in the equilibrium model affects the frequencies because of the elevation of the outer layers. However its perturbation is also important for the frequencies, as discussed in this paper. Although this subject has been already shown by \cite{Houdek10} using the equilibrium convection models and the TDC formalism based on the theory of \cite{Gough77b, Gough77a}, our study used the convection profiles obtained with the 3D simulations. As a first step, we limited ourselves to adiabatic oscillations and the effect of turbulent pressure. However, future works should consider nonadiabatic effects as well as the effect of convective backwarming. As for the latter, \cite{Trampedach13, Trampedach17} reported that the high temperature sensitivity of the opacity in the top of the convection zone causes warming by upflows of convection surpassing cooling by the downflows coupled with the non-linear nature of radiative transfer. The resultant net warming leads to the elevation of the outer layers as well as turbulent pressure. They also reported that the contribution of the backwarming has a similar magnitude as that of the turbulent pressure.

For the asteroseismology of solar-like stars, we need correct model frequencies. Since stellar ages are substantially affected by the surface effect, many studies have adopted the empirical relation based on the solar frequencies proposed by \cite{Kjeldsen08}. However \cite{Sonoi15} found that this solar-calibrated relation has difficulty in correcting the frequencies in different stellar models and at different evolutionary stages. Then, it may be preferable to find a method of the correction based on a strong physical approach. Especially, the convective effects both in the equilibrium state and perturbation and also the nonadiabatic effect may be important for this problem. Therefore, we will extend the work of \cite{Sonoi15} including these effects.

\begin{acknowledgements}
  T.S. has been supported by the ANR (Agence Nationale de la Recherche) program IDEE (Interaction Des \'Etoiles et des Exoplan\`etes) and CNES (Centre National d'\'Etudes Spatiales). H.G.L. acknowledges financial support by the Sonderforschungsbereich SFB 881 ``The Milky Way System'' (subproject A4) of the German Research Foundation (DFG). 
\end{acknowledgements}

\appendix
\section{Time-dependent convection formalism for nonlocal convection} 
\label{app:nonlocal_tdc}

Following \cite{Dupret06}, we introduce the way to adopt the results given by the 3D simulations of nonlocal convection to the time-dependent convection (TDC) formalism of \cite{Grigahcene05}.

\subsection{Hydrodynamical equations for local convection}
\label{sec:local_conv_equil}

The TDC formalism of \cite{Grigahcene05} originates from the one proposed by \cite{Unno67}. Later, \citeauthor{Unno67}'s formalism was developed for nonradial oscillations by \cite{Gabriel75}. The classical mixing length theory (MLT) of \cite{Bohm-Vitense58} is the description for convection in the hydrostatic equilibrium state. On the other hand, the TDC formalism includes variation of convection on the dynamical timescale. However, if we impose the stationary condition on the formalism, we can obtain consistent results with the MLT.

First, we derive the equation of convection in the equilibrium state. We thus decompose the physical variables in the hydrodynamical equations into the mean flow and convective fluctuations as $y=\overline{y}+\Delta y$ for the scalars and $\overrightarrow{v}=\overrightarrow{u}+\overrightarrow{V}$ for the velocity. In \citeauthor{Unno67}'s formalism, the convective fluctuation parts of the hydrodynamical equations of the continuity, movement, and energy conservation are given by
\begin{eqnarray}
  &&\nabla\cdot\overrightarrow{V}=0, \label{eq:cont} \\
  &&\overline{\rho}\frac{{\rm d}\overrightarrow{V}}{{\rm d}t}
  =\frac{\Delta\rho}{\overline{\rho}}\nabla\overline{p_{\rm tot}}
  -\nabla\Delta p_{\rm tot} - \rho\overrightarrow{V}\cdot\nabla\overrightarrow{u}
  -\Lambda\frac{\overline{\rho}\overrightarrow{V}}{\tau_c}, \label{eq:movement} \\
  &&\frac{\Delta(\rho T)}{\overline{\rho T}}\frac{{\rm d}\overline{s}}{{\rm d}t}
  +\overrightarrow{V}\cdot\nabla\overline{s}
  =-\frac{\omega_R\tau_c+1}{\tau_c}\Delta s, \label{eq:energy}
\end{eqnarray}
where the notations follow the definitions introduced in Sect. \ref{sec:adtdc} 
%%%\LEt{ of this paper? Please clarify here.} . 
of this paper.
To obtain the above equations, the following approximations have been made for the closure of Eqs. (\ref{eq:movement}) and (\ref{eq:energy}):
\begin{eqnarray}
  \Lambda\frac{\overline{\rho}\overrightarrow{V}}{\tau_c}
  &=&\frac{\Delta\rho}{\overline{\rho}}\nabla\cdot
  (\Delta\overline{\beta}_{\rm g}+\Delta\overline{\beta}_{\rm R}+\Delta\overline{\beta}_{\rm t}) 
  \nonumber \\
  &&-\nabla\cdot (\Delta\beta_{\rm g}+\Delta\beta_{\rm R}+\Delta\beta_{\rm t}), 
  \label{eq:move_close} \\
  \overline{\rho}\overline{T}\frac{\Delta s}{\tau_c}
  &=&-\overline{\rho}\overline{T}\overrightarrow{V}\cdot\nabla\overline{s}
  -\rho\epsilon_2+\overline{\rho\epsilon_2} \nonumber \\
  &&+(\rho T\nabla s)\cdot\overrightarrow{V}
  -\overline{(\rho T\nabla s)\cdot\overrightarrow{V}}, 
  \label{eq:energy_close1} \\
  \nabla\cdot\Delta\overrightarrow{F}_{\rm R}
  &=&-\omega_{\rm R}\Delta s\overline{\rho T}, 
  \label{eq:energy_close2} \\
  l&=&\alpha H_p=\alpha|{\rm d}r/{\rm d}\ln p_{\rm tot}|=|\overrightarrow{V}|\tau_c.
  \label{eq:lmix}
\end{eqnarray}
We adopt the Boussinesq approximation, in which the pressure fluctuations are neglected except in the equation of movement (Eq. \ref{eq:movement}) and the density fluctuations are neglected in the equation of continuity (Eq. \ref{eq:cont}). Besides, the spatial variation in the density is assumed to be much smaller than that in the convective velocity in Eq. (\ref{eq:cont}). The closure approximations (\ref{eq:move_close}), (\ref{eq:energy_close1}), and (\ref{eq:energy_close2}) follow the assumption that turbulent viscosity and thermal conductivity due to smaller eddies are expressed with the typical scale given by a representative convective element including them. Equation (\ref{eq:lmix}) is the usual closure equation of the MLT. Assuming constant coefficients and $\Lambda=8/3$, the above equations give the stationary solution consistent with the MLT:
\begin{eqnarray}
  &&\Gamma(\Gamma+1)={\cal A}(\nabla-\nabla_{\rm ad}), \label{eq:MLT1}\\
  &&\frac{9}{4}\Gamma^3+\Gamma^2+\Gamma={\cal A}(\nabla_{\rm rad}-\nabla_{\rm ad}), \label{eq:MLT2} \\
  &&F_c=\frac{\alpha^2c_p\rho T}{4}\sqrt{\frac{P_Tp_{\rm tot}}{2P_\rho\rho}}
  \left[\frac{\Gamma(\nabla-\nabla_{\rm ad})}{\Gamma+1}\right]^{3/2}, \label{eq:MLT_Fc} \\
  &&p_{\rm turb}=\frac{\alpha^2}{8}\frac{P_Tp_{\rm tot}}{2P_\rho}\frac{\Gamma}{\Gamma+1}(\nabla-\nabla_{\rm ad}),
  \label{eq:MLT_pturb}
\end{eqnarray}
where ${\cal A}=P_Tp_{\rm tot}/(2P_\rho\rho)[\kappa c_p\rho^2gl^2/(12acT^3p_{\rm tot})]^2$ and $\Gamma=(\omega_{\rm R}\tau_c)^{-1}$. As mentioned above, the Boussinesq approximation includes the neglect of the density fluctuations in Eq. (\ref{eq:cont}). However, this assumption is not valid in near-surface layers of solar-like stars since convective velocity can be comparable with sound speed of surrounding materials. Besides, the assumption that the spatial variation in the density is much smaller than that in the convective velocity is invalid in the deep part of a convection zone, where the surrounding structure is no longer homogeneous in the representative scale of convective eddies. However, this is a standard hypothesis made in most TDC approaches. Without such assumption, it is difficult to build a TDC formalism. Besides it is a consequence of the adoption of the MLT.

\subsection{Perturbative theory for local convection}
\label{app:local_conv_perturb}

To consider the behaviour of convection with the oscillations, we perturb the above formalism, Eqs. (\ref{eq:cont})--(\ref{eq:energy}), which allows us to evaluate the perturbation of correlated quantities of the convective fluctuations. However the closure described above is crude, and many complex physical processes are neglected including the whole cascade of energy. Then, uncertainty cannot be avoided when perturbing the closure terms. Because of such uncertainties, the unphysical, short wavelength oscillations appear in the eigenfunctions of the differential equations for the oscillations. To deal with this problem, \cite{Grigahcene05} proposed to introduce a free complex parameter $\beta$ in the perturbation of the thermal closure equations:
\begin{eqnarray}
  \delta\left(\frac{\Delta s}{\tau_c}\right)=\frac{\Delta s}{\tau_c}
  \left[(1+\beta\sigma\tau_c)\frac{\delta\Delta s}{\Delta s}
    -\frac{\delta\tau_c}{\tau_c}\right].
  \label{eq:beta}
\end{eqnarray}
Introducing this parameter leads to phase lags between the oscillations and the way the turbulence cascade adapts to them.

Then, we search for the solutions of the perturbed convective fluctuation equations of the form $\delta(\Delta X)=\delta(\Delta X)_{\overrightarrow{k}}e^{i\overrightarrow{k}\cdot\overrightarrow{r}}e^{i\sigma t}$, assuming constant coefficients within the coarse grain, which is much larger than most of convective eddies but much smaller than the scale of the perturbation wavelength. Next, we integrate these particular solutions over all values of $k_\theta$ and $k_\phi$ so that $k^2_\theta+k^2_\phi=Ak^2_r$, keeping $A$ constant and that every direction of the horizontal component of $\overrightarrow{k}$ has the same probability. The value of $A$ is the free parameter, given by Eq. (\ref{eq:A}) based on the 3D simulation in this study. We have to introduce this distribution of $\overrightarrow{k}$ values to obtain an expression for the perturbation of the Reynolds tensor which allows the proper separation of the variables in terms of spherical harmonics in the equation of motion. Finally, the obtained values of the perturbation of the correlated values are implemented into the differential equations of the oscillations.

\subsection{Procedure for taking equilibrium values given by 3D simulation into account}
3D hydrodynamic simulations \cite[e.g.][]{Stein91, Stein98, Rosenthal99, Yang07, Piau14} provide much more realistic profiles of the convection zones than with the MLT. Here we discuss how to extend the above formalism to the nonlocal case following \cite{Dupret06}.

As discussed in Sect. \ref{sec:local_conv_equil}, most of the uncertainties are included in the closure terms (Eqs. \ref{eq:move_close}--\ref{eq:lmix}). Then, we modify these terms introducing a free function varying with depth, $\Omega$, which has the same meaning as in the formalism of \cite{Canuto91}. It can be assumed to be a function of the convective efficiency $\Gamma$ following \citeauthor{Canuto91}. We also set the usual mixing length $\alpha$ as an additional free function varying with the depth or $\Gamma$. More precisely, we multiply the left hand side of Eq. (\ref{eq:move_close}) by $\Omega(\Gamma)$ and the left hand side of Eqs. (\ref{eq:energy_close1}) and (\ref{eq:energy_close2}) by $1/\Omega(\Gamma)$.
Then, Eqs. (\ref{eq:movement}) and (\ref{eq:energy}) become
\begin{eqnarray}
  &&\overline{\rho}\frac{{\rm d}\overrightarrow{V}}{{\rm d}t}
  = \frac{\Delta\rho}{\overline{\rho}}\nabla\overline{p_{\rm tot}} 
  - \nabla\Delta p_{\rm tot} 
  - \rho\overrightarrow{V}\cdot\nabla\overrightarrow{u} 
  - \Omega(\Gamma)\Lambda\frac{\overline{\rho}\overrightarrow{V}}{\tau_c}, 
  \label{eq:movement_mod} \\
  &&\frac{\Delta(\rho T)}{\overline{\rho T}}\frac{{\rm d}\overline{s}}{{\rm d}t}
  + \frac{{\rm d}\Delta s}{{\rm d}t} + \overrightarrow{V}\cdot\nabla\overline{s}
  = - \frac{\omega_R\tau_c+1}{\Omega(\Gamma)\tau_c}\Delta s.
  \label{eq:energy_mod}
\end{eqnarray}
In the stationary case, these new equations have a form similar to the old ones (Eqs. \ref{eq:movement} and \ref{eq:energy}). Equation (\ref{eq:MLT1}) remains unchanged, giving the same meaning to $\Gamma$ as in the previous case. Equation (\ref{eq:MLT_pturb}) is still verified (with varying $\alpha$), but Eqs. (\ref{eq:MLT2}) and (\ref{eq:MLT_Fc}) are slightly modified:
\begin{equation}
  \frac{9}{4}\Omega(\Gamma)\Gamma^3+\Gamma^2+\Gamma={\cal A}(\nabla_{\rm rad}-\nabla_{\rm ad}),
  \label{eq:MLT2_nl}
\end{equation}
\begin{equation}
  F_{\rm c,l}
  =\frac{\Omega(\Gamma)\alpha^2c_p\rho T}{4}\sqrt{\frac{P_Tp_{\rm tot}}{2P_\rho\rho}}
  \left[\frac{\Gamma(\nabla-\nabla_{\rm ad})}{\Gamma+1}\right]^{3/2}, 
  \label{eq:Fc_om}
\end{equation}
\begin{equation}
  p_{\rm turb,l}
  =\frac{\alpha^2}{8}\frac{P_Tp_{\rm tot}}{2P_\rho}\frac{\Gamma}{\Gamma+1}(\nabla-\nabla_{\rm ad}). 
  \tag{\ref{eq:MLT_pturb}b}
\end{equation}
By adjusting $\Omega$ and $\alpha$, we can fit these equations to the results given by 3D simulations in combination with Eqs. (\ref{eq:ptnl}) and (\ref{eq:fcnl}).
%%%\LEt{ avoid starting sentence with symbols.}  
The quantities
$F_{\rm c,nl}$, $p_{\rm turb,nl}$, $(\nabla-\nabla_{\rm ad})$ and other thermodynamic quantities are deduced from the 3D simulations, and we take appropriate horizontal and time averages. Using Eqs. (\ref{eq:ptnl}) and (\ref{eq:fcnl}), the local counterparts of turbulent pressure and convective flux, $p_{\rm turb,l}$ and $F_{\rm c,l}$, are obtained based on $p_{\rm turb,nl}$ and $F_{\rm c,nl}$. Using these local counterparts, we obtain appropriate values of $\Omega$ and $\alpha$ at each location with Eqs. (\ref{eq:Fc_om}) and (\ref{eq:MLT_pturb}b), respectively.

To generalize the perturbative theory presented in Sect. \ref{app:local_conv_perturb}, we replace the equations of movement and energy conservation for the local case, (\ref{eq:movement}) and (\ref{eq:energy}), with the ones for the 3D case, (\ref{eq:movement_mod}) and (\ref{eq:energy_mod}). We follow the same procedure as in Sect. \ref{app:local_conv_perturb}. Assuming again constant coefficient and searching for solutions in the form of plane waves, we obtain the new expressions for the perturbed local convective quantities such as the convective flux and turbulent pressure.

The main uncertainties in this approach appear in the way to perturb $\Omega$ and $\alpha$. The free parameter $\beta$ introduced in Eq. (\ref{eq:beta}) is also somehow related to these uncertainties. At present, we have no theoretical prescription how to perturb $\Omega$ and $\alpha$, and then we neglect their perturbations. However there is no reason to expect them to be small, and we should not be too optimistic when using this new perturbative treatment.

Here, we do not discuss the derivation which 
%%%are \LEt{ derivations are or derivation is.} 
is
very similar to those of \cite{Grigahcene05}. The final results of the perturbation of the radial components of the local convective velocities and convective flux are given as Eqs. (\ref{eq:dVr_nad}) and (\ref{eq:dfc}). They are not so different from the former expressions \citep[Eqs. 12 and 18 in][]{Grigahcene05}:
\begin{eqnarray}
  \frac{\overline{V_r\delta V_r}}{\overline{V^2_r}}
  &=&\frac{1}{B+[(i\Omega+\beta)\sigma\tau_c+1]D}
  \nonumber\\
  &&\cdot\left\{-\frac{\delta c_p}{c_p}-\frac{\delta Q}{Q} - \frac{\delta\rho}{\rho}
  +\totdif{\delta p_{\rm tot}}{p_{\rm tot}} - \totdif{\xi_r}{r}\right. \nonumber\\
  &&-i\Omega\sigma\tau_cD\frac{Q+1}{Q}\frac{\delta s}{c_p}
  +C\left[\totdif{\delta s}{s}-\totdif{\xi_r}{r}\right] \nonumber\\
  &&-\frac{A}{A+1}\frac{i\sigma\tau_c}{\Omega\Lambda}
  \left[\totdif{\xi_r}{r}+\frac{1}{A}\frac{\xi_r}{r}-\frac{\ell(\ell+1)}{2A}\frac{\xi_h}{r}\right]
  \nonumber\\
  &&-\omega_R\tau_cD\left(3\frac{\delta T}{T}-\frac{\delta c_p}{c_p}
  -\frac{\delta\kappa}{\kappa}-2\frac{\delta\rho}{\rho}\right) \nonumber\\
  &&\left.+[(i\Omega+\beta)\sigma\tau_c+3\omega_R\tau_c+2]D\frac{\delta l}{l}\right\},
  \label{eq:dVr_nad}
\end{eqnarray}
\begin{eqnarray}
  \frac{\delta F_{\rm c,l}}{F_{\rm c,l}}&=&\frac{\delta\rho}{\rho}+\frac{\delta T}{T}
  -i\Omega\sigma\tau_cD\frac{Q+1}{Q}\frac{\delta s}{c_p}
  +C\left[\frac{{\rm d}\delta s}{{\rm d}s}-\totdif{\xi_r}{r}\right] \nonumber\\
  &&-\omega_R\tau_cD\left(\frac{\delta T}{T}-\frac{\delta c_p}{c_p}
  -\frac{\delta\kappa}{\kappa}-2\frac{\delta\rho}{\rho}\right) \nonumber\\
  &&+[(i\Omega+\beta)\sigma\tau_c+2\omega_R\tau_c+1]D\frac{\overline{V\delta V_r}}{\overline{V^2_r}}
  \nonumber\\
  &&+(2\omega_R\tau_c+1)D\frac{\delta l}{l}.
  \label{eq:dfc}
\end{eqnarray}
We note that Eq. (\ref{eq:dVr_nad}) becomes Eq. (\ref{eq:dVr}) for adiabatic radial oscillations.

On the other hand, the perturbation of the nonlocal turbulent pressure and convective flux is obtained by solving the eigenvalue problem of the differential equations of oscillations combining the perturbed equations of (\ref{eq:ptnl}) and (\ref{eq:fcnl}).

\section{Recasting the expression of turbulent pressure perturbation}
\label{app:transf_dpt}

Here, we describe the procedure for recasting the expression of turbulent pressure perturbation, which is required for the discussion in Sect. \ref{sec:main_eff}. We begin with Eqs. (\ref{eq:dpturb}) and (\ref{eq:dVr}) and aim to express them as the linear combination of the thermal pressure perturbation, $\delta p_{\rm th}$, and the displacement, $\xi$.

For the perturbation of the mixing length, we adopt the expression,
\begin{eqnarray}
  \frac{\delta l}{l}=\frac{1}{1+(\sigma\tau_c)^2}\frac{\delta H_p}{H_p},
  \label{eq:dll}
\end{eqnarray}
where the perturbation of the pressure scale height is described by
\begin{eqnarray}
  \frac{\delta H_p}{H_p}=\frac{\delta p_{\rm tot}}{p_{\rm tot}}-\totdif{\delta p_{\rm tot}}{p_{\rm tot}}+\totdif{\xi}{r}.
\end{eqnarray}
To cancel the term ${\rm d}\delta p_{\rm tot}/{\rm d}p_{\rm tot}$, we adopt the perturbed equation of movement \citep[Eq. D.3 in ][]{Grigahcene05}, neglecting the perturbation of the divergence of the Reynolds tensor:
\begin{eqnarray}
  \totdif{\delta p_{\rm tot}}{p_{\rm tot}}=-\frac{\sigma^2r}{g}\frac{\xi}{r}+\frac{1}{g}\totdif{\delta\Phi}{r}
  +\frac{\delta\rho}{\rho}+\frac{2A-1}{A}\frac{p_{\rm turb,l}}{\rho gr}\totdif{\xi}{r},
  \label{eq:pulsmov}
\end{eqnarray} 
and the perturbed equation of continuity \citep[Eq. D.1 in][]{Grigahcene05},
\begin{eqnarray}
  \frac{\delta\rho}{\rho}+\frac{1}{r^2}\totdif{}{r}\left(r^2\xi\right)=0.
  \label{eq:pulscont}
\end{eqnarray}
For radial oscillation, the Poisson equation becomes
\begin{eqnarray}
  \frac{1}{g}\totdif{\delta\Phi}{r}=\totdif{\xi}{r}.
  \label{eq:poisson}
\end{eqnarray}
With the adiabatic condition $\delta \rho/\rho=\delta p_{\rm th}/p_{\rm th}/\Gamma_1$, we can express Eq. (\ref{eq:dVr}) as a linear combination of $\delta p_{\rm th}$, $\xi$ and $\delta p_{\rm turb,l}$. We categorize the terms in Eq. (\ref{eq:dVr}) into four parts as follows. The first part is the term which stems from the advection term in the equation of movement:
\begin{eqnarray}
  -\frac{A}{A+1}\frac{i\sigma\tau_c}{\Omega\Lambda}
  \left(\totdif{\xi}{r}+\frac{1}{A}\frac{\xi}{r}\right)=
  \Pi_1\frac{\delta p_{\rm th}}{p_{\rm th}}+\Xi_1\frac{\xi}{r}.
  \label{eq:part1}
\end{eqnarray}
The second part is the perturbation of the mixing length:
\begin{eqnarray}
  (D+1)\frac{\delta l}{l}=
  \tau_2\frac{\delta p_{\rm turb,l}}{p_{\rm turb,l}}+
  \Pi_2\frac{\delta p_{\rm th}}{p_{\rm th}}+\Xi_2\frac{\xi}{r}.
  \label{eq:part2}
\end{eqnarray}
The third part is the perturbation of the inverse of the radiative cooling timescale of convection eddies:
\begin{eqnarray}
  -\omega_R\tau_cD\frac{\delta\omega_R}{\omega_R}=
  \tau_3\frac{\delta p_{\rm turb,l}}{p_{\rm turb,l}}+\Pi_3\frac{\delta p_{\rm th}}{p_{\rm th}}+
  \Xi_3\frac{\xi}{r}.
  \label{eq:part3}
\end{eqnarray}
The last part corresponds to the remaining terms:
\begin{eqnarray}
  -\frac{\delta c_p}{c_p}-\frac{\delta Q}{Q} - \frac{\delta\rho}{\rho} + \totdif{\delta p_{\rm tot}}{p_{\rm tot}}
  -(C+1)\totdif{\xi}{r}=
  \Pi_4\frac{\delta p_{\rm th}}{p_{\rm th}}+\Xi_4\frac{\xi}{r}.
  \label{eq:part4}
\end{eqnarray}
Finally, we obtain the expression of $\delta p_{\rm turb,l}/p_{\rm tot}$ as Eq. (\ref{eq:dpturb_local}). The coefficient of $\delta p_{\rm th}/p_{\rm tot}$ ($\Pi$) consists in $\Pi_i$'s ($i=1,2,3,4$) as shown in Eq. (\ref{eq:Pi}). The expression of $\Pi_i$'s is
\begin{eqnarray}
  \Pi_1&=&\frac{A}{A+1}\frac{i\sigma\tau_c}{\Omega\Lambda}\frac{1}{\Gamma_1}, 
  \label{eq:pi1}\\
  \Pi_2&=&\frac{D+1}{1+(\sigma\tau_c)^2}
  \left[\frac{p_{\rm th}}{p_{\rm tot}}+\frac{2A-1}{A}\frac{p_{\rm turb,l}}{\rho gr}-\frac{1}{\Gamma_1}\right], \\
  \Pi_3&=&-\omega_R\tau_cD
  \left(3\nabla_{\rm ad}-c_{p,{\rm ad}}-\kappa_{\rm ad}-\frac{2}{\Gamma_1}\right.\nonumber\\
  &&\left.-\frac{2}{1+(\sigma\tau_c)^2}
  \left[\frac{p_{\rm th}}{p_{\rm tot}}+\frac{2A-1}{A}\frac{p_{\rm turb,l}}{\rho gr}-\frac{1}{\Gamma_1}\right]\right), \\
  \Pi_4&=&-c_{p,{\rm ad}}-Q_{\rm ad}+\frac{C}{\Gamma_1}+\frac{2A-1}{A}\frac{p_{\rm turb,l}}{\rho gr}.
  \label{eq:pi4}
\end{eqnarray}
On the other hand, the 
%%%coefficient 
coefficients
of $\xi/R$, $\Xi$, are
%%%\LEt{ coefficient is or coefficients are.} 
\begin{eqnarray}
  \Xi_1&=&\frac{A}{A+1}\frac{i\sigma\tau_c}{\Omega\Lambda}\frac{2A-1}{A}, 
  \label{eq:xi1}\\
  \Xi_2&=&\frac{D+1}{1+(\sigma\tau_c)^2}
  \left[\frac{\sigma^2r}{g}+2\frac{2A-1}{A}\frac{p_{\rm turb,l}}{\rho gr}\right], \\
  \Xi_3&=&\frac{2\omega_R\tau_cD}{1+(\sigma\tau_c)^2}
  \left[\frac{\sigma^2r}{g}+2\frac{2A-1}{A}\frac{p_{\rm turb,l}}{\rho gr}\right], \\
  \Xi_4&=&-\frac{\sigma^2r}{g}-2\frac{2A-1}{A}\frac{p_{\rm turb,l}}{\rho gr}+2C. 
  \label{eq:xi4}
\end{eqnarray}

%-------------------------------------------------------------------
\bibliographystyle{aa}
\bibliography{sonoi}

\end{document}